\documentclass[utf8]{frontiersSCNS} % for Science, Engineering and Humanities and Social Sciences articles
\usepackage{url,hyperref,lineno,microtype,subcaption}
\usepackage[onehalfspacing]{setspace}
\usepackage{aas_macros}
\usepackage{siunitx}

\newcommand{\Hb}{\rm H\beta}
\newcommand{\Ha}{\rm H\alpha}

\def\keyFont{\fontsize{8}{11}\helveticabold }
\def\firstAuthorLast{Congiu {et~al.}} %use et al only if is more than 1 author
\def\Authors{E. Congiu\,$^{1,2,*}$, M. Contini\,$^{3}$, S. Ciroi\,$^{1}$, V. Cracco\,$^{1}$, F. Di Mille\,$^4$, M. Berton\,$^{1,2}$,\\
M. Frezzato\,$^{1}$, G. La Mura\,$^{1}$, P. Rafanelli\,$^{1}$}
\def\Address{$^{1}$ Dipartimento di Fisica e Astronomia ”G. Galilei”, Universit\`a di Padova, Vicolo dell’Osservatorio 3, I-35122 Padova, Italy;\\
$^2$ INAF-Osservatorio Astronomico di Brera, via E. Bianchi 46, I-23807, Merate (LC), Italy;\\
$^3$ School of Physics and Astronomy, Tel Aviv University, Tel Aviv 69978, Israel;\\
$^4$ Las Campanas Observatory - Carnegie Institution of Washington, Colina el Pino Casilla 601, La Serena, Chile.}
\def\corrAuthor{E. Congiu}

\def\corrEmail{enrico.congiu@phd.unipd.it}

\begin{document}
\onecolumn
\firstpage{1}

\title[ENLR in Seyfert galaxies]{Extended narrow-line region in Seyfert galaxies} 

\author[\firstAuthorLast ]{\Authors} %This field will be automatically populated
\address{} %This field will be automatically populated
\correspondance{} %This field will be automatically populated

\extraAuth{}% If there are more than 1 corresponding author, comment this line and uncomment the next one.
%\extraAuth{corresponding Author2 \\ Laboratory X2, Institute X2, Department X2, Organization X2, Street X2, City X2 , State XX2 (only USA, Canada and Australia), Zip Code2, X2 Country X2, email2@uni2.edu}

\maketitle

\begin{abstract}

%%% Leave the Abstract empty if your article does not require one, please see the Summary Table for full details.
We present our recent results about the extended narrow-line region (ENLR) of two nearby Seyfert 2 galaxies (IC\,5063 and NGC\,7212) obtained by modelling the observed line profiles and spectra with composite models (photoionization+shocks) in the different regions surrounding the AGN.
Then, we compare the Seyfert 2 ENLRs with the very extended one recently discovered in the narrow-line Seyfert 1 (NLS1) galaxy Mrk\,783.
We have found several evidences of interaction between the ISM of the galaxies and their radio jets, such as a) the contribution of shocks in ionizing the high velocity gas, b) the complex kinematics showed by the profile of the emission lines, c) the high fragmentation of matter, etc.
The results suggest that the ENLR of IC\,5063 have a hollow bi-conical shape, with one edge aligned to the galaxy disk, which may cause some kind of dependence on velocity  of the ionization parameter.
Regarding the Mrk 783 properties, it is found that the extension of the optical emission is almost twice the size of the radio one and it seems  
due to the AGN activity, although there is contamination by star formation around $12$ arcsec from the nucleus.
Diagnostic diagrams excluded the contribution of star formation in IC\,5063 and NGC\,7212, while the shock contribution was used to explain the spectra emitted by their high velocity gas.

\tiny
 \keyFont{ \section{Keywords:} active galactic nuclei, Seyfert galaxies, emission lines, extended narrow-line region, IC\,5063, NGC\,7212, Mrk\,783.} %All article types: you may provide up to 8 keywords; at least 5 are mandatory.
\end{abstract}

\section{Introduction}
\label{sec:1}

The extended narrow-line region (ENLR) is  one of the most intriguing structures which characterize active galactic nuclei (AGN).
Predicted by the unified model \citep{Antonucci85,Antonucci93}, the ENLR is a region of highly ionized gas which is probably produced by the ionizing radiation escaping the AGN along the axis of the dusty torus.
It is characterized by strong and narrow optical emission lines, both permitted and forbidden, similar to those emitted from the narrow-line region (NLR)\footnote{We consider the ENLR the natural extension of the NLR over $\sim 1$ kpc}, which usually can be traced up to few kiloparsecs from the nucleus.
However, in  some cases,  the strongest lines (e.g. [O III]$\lambda5007$) can be traced up to $\sim20$\,kpc or more \citep{Mulchaey94,Schmitt03}. 

The ENLR is often characterized by a conical or bi-conical shape whose apexes are pointing towards the active nucleus.
In these cases it can be referred to as \emph{ionization cones}.
They are usually observed in nearby Seyfert 2 galaxies, but only a few tens of them are known and well studied \citep[$\sim 50$ at $z<0.05$,][]{Netzer15}.
ENLRs are also observed in Seyfert 1 galaxies, but because of the origin of the ionizing radiation, they are expected to be smaller and halo-like (not conical) \citep{Evans93,Schmitt03,Schmitt03b}.
Although this is true for most Seyfert 1, some exceptions, such as NGC\,4151, show a conical ENLR \citep{Pogge89}.
\citet{Mulchaey96}, showed that the ENLR shape is not only determined by the orientation of the AGN, but it strongly depends on the gas distribution  throughout the galaxy.
The ENLR gas is usually considered as the host galaxy gas ionized by the AGN, but in some objects, e.g. NGC\,4388 \citep{Ciroi03}, Mrk 315 \citep{Ciroi05}, NGC 7212 \citep{Cracco11}, merging with small gas rich galaxies could provide a new gas supply.
The fast outflows observed in several galaxies \citep[e.g.][]{Baldwin87,Morganti15,Dasyra15} could also be an important source of gas.

A peculiar characteristic of the ENLR consists in its complex kinematics.
Emission lines with multiple peaks and asymmetries are observed in several objects \citep[e.g.][]{Dietrich98,Morganti07,Ozaki09,Cracco11,Congiu17b} and high resolution images \citep{Schmitt03} indicate that the gas is  concentrated in filaments and cloud substructures.
Mergers and outflows could be responsible for this kind of kinematics, likewise the interaction between the galaxy interstellar medium (ISM) and the AGN radio jet.
The ionization cone axis is often aligned with that of the galaxy radio emission \citep{Wilson94,Nagar99,Schmitt03,Schmitt03b} and strong outflows in all the gas phases are observed near radio hotspots \citep{Morganti98,Morganti07,Morganti15}.

To have a better insight of the ENLR gas properties, we have studied high resolution spectra of two nearby Seyfert 2 galaxies with ENLR: IC\,5063 and NGC\,7212 \citep{Congiu17b}.
In  sections 2 and 3 we summarize the main results of that work, while in section 4 we compare them with the ENLR in Mrk 783, a narrow-line Seyfert 1 (NLS1) galaxy with a recently discovered extended radio emission \citep{Congiu17}.

\section{Sample and observations}
\label{sec:2}

IC\,5063 and NGC\,7212 , two Seyfert 2 galaxies with well known ENLR were the
targets of our observations.
IC\,5063 is a lenticular galaxy ($z=0.01135$) characterized by a complex system of dust lanes, aligned with the major axis of the galaxy and by a very bright radio source \citep{Morganti98}.
The galaxy shows fast outflows in all the gas phases, from cold molecular gas to warm ISM \citep{Morganti98,Morganti07,Tadhunter14,Dasyra15,Morganti15}.
Its ENLR was first discovered by \citet{Colina91}.
NGC\,7212 is a spiral galaxy ($z=0.02663$) interacting with other two galaxies of its group \citep[e.g.][]{Wasilewski81}.
The ENLR was first discovered in polarized light by \citet{Tran95} then \citet{Falcke98} found an extended NLR also in non-polarized light. \citet{Cracco11} confirmed the presence of ionization cones on both sides of the galaxy.
The targets were observed with the MagE (Magellan echellette) spectrograph of the Magellan telescopes of Las Campanas Observatory.
The instrument covers the whole optical spectrum ($3100-10000\,\si{\angstrom}$) with a resolution R $\sim 8000$ \citep{Marshall08}.
All the data needed for reduction and calibration process were also acquired.
The data reduction process is described in details in \citet{Congiu17b}.

\section{Gas properties as a function of velocity}
\label{sec:3}

We analyzed the high resolution spectra of the ENLR because they allow to  understand the complex profiles of the emission lines. 
Moreover, \citep[see][]{Ozaki09}  it is possible to study the properties of the gas clouds as a function of velocity.
\citet{Ozaki09} divided the profile of few important emission lines in velocity bins, according to the peculiarities of the profile, and he analyzed the gas properties in these bins by means of line ratios and photo-ionization simulations.
We applied his method to our spectra of IC\,5063 and NGC\,7212 increasing the number of studied lines, uniforming the width of the velocity bins to $100$ km s$^{-1}$ and reproducing the observed spectra with composite models \footnote{SUMA models consider both photo-ionization and shocks at the same time to reproduce the line ratios of the analyzed spectrum} calculated by the code SUMA \citep{Contini02}.
We divided the two-dimensional spectra in several regions, to study the behavior of the gas properties as a function of the distance from the nucleus. 
The complete data analysis is shown in \citet{Congiu17b}.
With this method we were able to produce diagnostic diagrams and several plots of the principal gas properties as a function of velocity for each examined region.

The analysis of the emission line profiles confirmed that the kinematics of the analyzed gas is complex and that the emission lines are often composed by several components which strongly vary from region to region while the profiles observed in the same region are mostly constant. 
The variations in some cases occur because the physical conditions and the ion fractional abundances change moving from region to region.
To study the ionization mechanism of the gas we compared the result of four different diagnostic diagrams.
Three of them are the classical BPT diagrams from \citet{Baldwin81} and
\citet{Veilleux87} ($\log([\rm O\,III]/\Hb$ vs $\log([\rm N\,II]/\Ha)$, $\log([\rm S\,II]/\Ha)$, $\log([\rm O\,I]/\Ha)$) while the last one is a combination of the previous ratios with [O II]$\lambda3727$/[O III]$\lambda5007$, which is able to discern between photo-ionization by a power-law, star formation or ionization by shocks \citep[$\Delta E$ diagram,][]{Baldwin81}. 
All the diagnostic diagrams show that the main ionization mechanism of the gas is photo-ionization by the AGN.
There is no substantial contamination by star formation in any of our spectra.
However, the high velocity gas often show low $\log([\rm O III]/\Hb)$ and high ratios of low ionization lines to $\Ha$, shifting their points towards the LINER region of the BPT diagrams and the shock region of the $\Delta E$ diagram \citep[Fig\,6][]{Congiu17b}.
This should be interpreted as a strong contribution of shocks to the ionization of high velocity gas.
Shocks are most likely created by the interaction of the AGN jets with the galaxy ISM.
The results of composite models confirm this hypothesis.
The ratios of the low ionization lines to $\Hb$ in the high velocity bins can be reproduced accurately only adopting shocks with velocities comparable with the velocity of the gas, while the spectra of the low velocity bins can be reproduced using radiation dominated models.

The models also show that a wide range of cloud geometrical thicknesses is needed to reproduce the spectra.  
This is the result of cloud fragmentation in the turbulent regime created by the shocks which originate from the interaction of the jet with the ISM.
According to \citet{Roche16}, this kind of interaction should increase the temperature of the gas. 
Indeed, this is observed in the nuclear regions of NGC\,7212, the only regions of this object in which we were able to measure this quantity.
Using the total flux of the lines, the galaxy show slightly higher temperatures than expected \citep[$\sim 15000$ K with respect to the expected $10000$ K][]{OsterbrockAGN}.
Adopting composite models, the temperature is maintained at $10000$--$20000$ K by diffuse secondary radiation \citep[e.g. Fig. 10][]{Contini12} in a large region within the clouds.  
The temperatures evaluated in IC\,5063 are closer to $10000$ K.

Another interesting result is the behavior of the ionization parameter $U$ as a function of velocity in IC\,5063 \citep[Fig.\,8][]{Congiu17b}.
$U$ (defined as the number of ionizing photons reaching the cloud per number 
of electrons in the gas) is usually used to estimate the ionization degree of the gas.
We measured $U$ for each velocity bin using the \citet{Penston90} relation, 
which links $U$ to the [O II]$\lambda3727$/[O III]$\lambda5007$ ratio.
$U$ is not expected to depend on the velocity of the gas, in fact for NGC\,7212 it can be considered almost constant in all the velocity bins of the same region.
However, for IC\,5063 a clear dependence is noticed, similar to what observed by \citet{Ozaki09} in NGC\,1068.
Following his model, a possible explanation could be the orientation 
of the ionization cones with respect to the line of sight.
This suggests that the ENLR has a hollow bi-conical shape with one edge  lying on the galactic disk.
This part of the cones is ionized by a partly absorbed flux leading to a lower ionization degree of the gas with respect to the side of the cone reached by the unabsorbed continuum.

Finally, by the detailed modelling of the spectra we obtained the metallicity of the gas.
Previous results \citep[e.g.][]{Contini17} show that most AGN and HII regions can be reproduced with solar metallicities.
However, we tried to reproduce the spectra with several values of metallicity.
The best results for all the regions and all the bins of the ionization cones could be obtained using abundances close to the \citet{Grevesse98} solar ones in agreement with the results previously mentioned.
Comparing our results with those obtained by modelling spectra observed outside the ionization cone, could be a way to study the origin of the gas.
If a difference in metallicity is detected, the origin of the ENLR gas might be different from that of the galaxy.

\section{Comparison with NLS1: Mrk 783}
\label{sec:mrk783}

\begin{figure}[h!]
\begin{center}
\includegraphics[width=\textwidth]{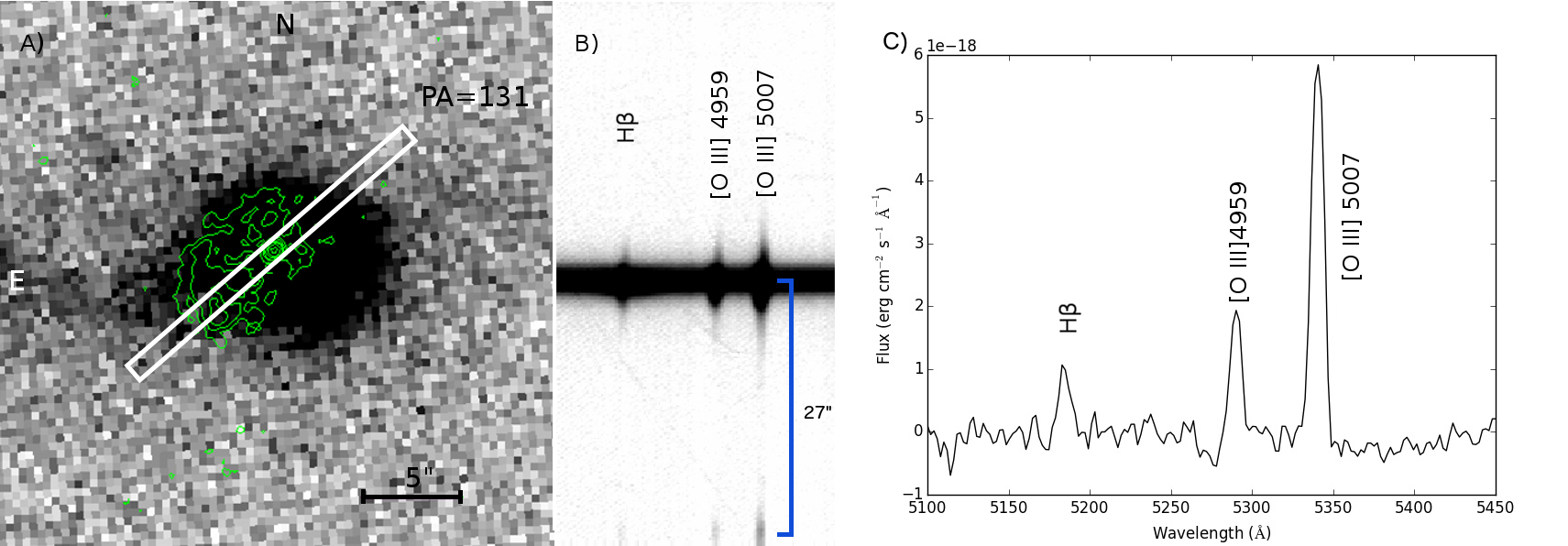}% This is a *.jpg file
\end{center}
\caption{From left to right: \textbf{A)} SDSS g-band image of Mrk 783 with the contours of the emission at $5\,\si{GHz}$. The contours are $3, 6,12,24,48,96$ and $192 \times\sigma$ ($\sigma = 11\,\si{\micro Jy.beam^{-1}}$). The beam size is $0.45\times0.40\,\si{arcsec}$ and the scale is $1.3\,\si{kpc.arcsec^{-1}}$. The position of the slit used for the observations with the Magellan telescope is shown.
The galaxy is characterized by a radio emission with a maximum extension of $\sim 5\,\si{arcsec}$ on the south-east side of the nucleus \citep{Congiu17}.
\textbf{B)} A small region of the LDSS3 spectra of Mrk 783 containing three emission lines: H$\beta$, [O III]$\lambda4959$ and [O III]$\lambda5007$. The data were acquired with a total exposure time of $1$ hour, with a seeing of
$0.8\,\si{arcsec}$ and the slit oriented as in panel A. The spatial resolution is $0.189\,\si{arcsec.px^{-1}}$ and the dispersion is $\sim 2\,\si{\angstrom.px^{-1}}$ .
\textbf{C)} Plot of the spectrum in panel B at $27\,\si{arcsec}$ from the galactic nucleus.}
\label{fig:mrk783}
\end{figure}

The ENLR is often associated with extended radio emission, both in Seyfert 1 and Seyfert 2 galaxies.
An extended radio emission with very steep in band spectral index was discovered in the NLS1 Mrk 783 \citep{Congiu17}. 
The optical follow-up of the source (Congiu et al., in preparation) shows an extended optical emission aligned to the radio axis.
The observed spectrum has a spatial resolution of $0.189$ arcsec px$^{-1}$ and a dispersion of $\sim 2$ $\si{\angstrom}$ px$^{-1}$.
Fig.\,\ref{fig:mrk783} shows the radio map of the object and the H$\beta$ region of the acquired spectrum.

\begin{figure}[h!]
\begin{center}
\includegraphics[height=7.5cm]{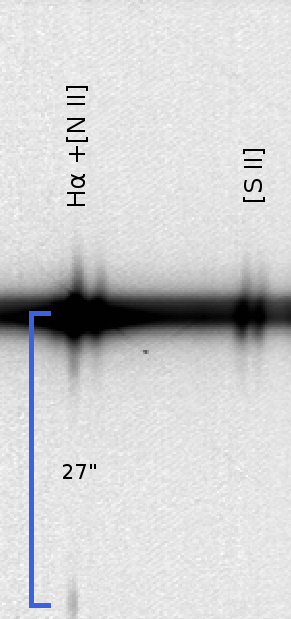} \quad% This is a *.jpg file
\includegraphics[width=0.6\textwidth]{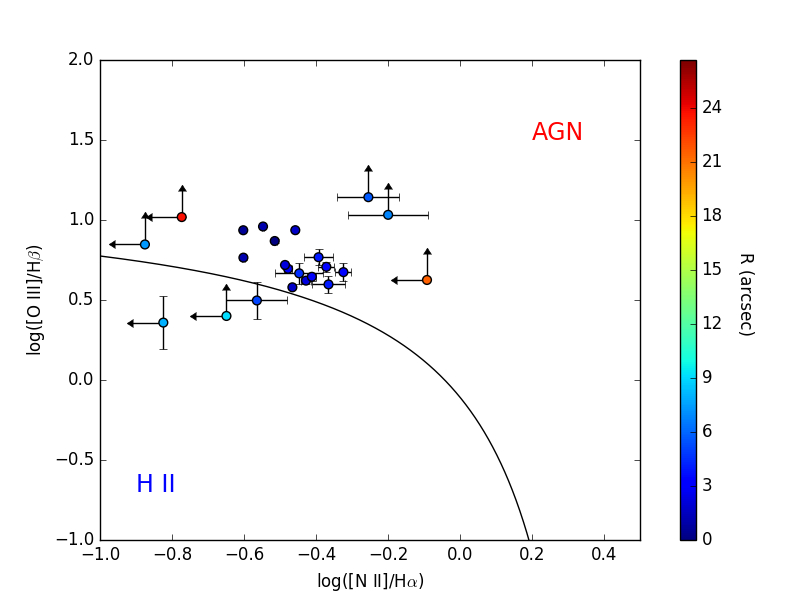}% This is a *.jpg file
\end{center}
\caption{\textbf{Left:} H$\alpha$ region of the same spectrum in Fig.\,\ref{fig:mrk783}, panel B. \textbf{Right:} Preliminary diagnostic diagram of the ENLR of Mrk 783. The colorbar show the distance of the bin from the nucleus. The black line dividing the plot in two regions is the relation from \citet{Kewley06} for extreme starburst. The arrows show the limit derived where we could not measure all the emission line.}
\label{fig:diag783}
\end{figure}

The [O III] and H$\beta$ lines were tracked  up  to $\sim 27$ arcsec from the nucleus of the galaxy (Fig.\,\ref{fig:mrk783}, panel B), which corresponds to a projected dimension of $\sim 35$ kpc. 
This makes this ENLR one of the most extended discovered so far.
Also H$\alpha$ can be traced up to the same distance, while none of the other low ionization lines (e.g. [N II]$\lambda\lambda6548,6584$, [S II]$\lambda\lambda6717,6731$) can be detected in the most extended emission region (Fig.\,\ref{fig:diag783} left panel). 
Due to the sharp end of the emission at the edge of the slit, it seems that the ENLR can extend even further. 
For comparison, the MagE spectra of IC\,5063 and NGC\,7212 do not cover the whole extension of the ionization cones, however \citet{Morganti07} measured a maximum extension of $\sim 3.8$ kpc in IC\,5063 and \citet{Cracco11} measured $\sim 4$ kpc in NGC\,7212.
An ENLR with a similar size is observed in NGC\,5252, which has a maximum extension of $\sim 33$ kpc \citep{Tadhunter89}.

It is also worth noting that the optical emission is far more extended with respect to the radio one and that it is observed only on the south-east side of the nucleus, while in our other sources the ionization cones are observed on both sides of the nucleus.
This is probably the consequence of a strong extinction or of a lack of gas in the north-west side of the nucleus.
There is also a region between the nucleus and the most extended emission, from $\sim12$ to $\sim22\,\si{arcsec} (\sim 16$--$28\,\si{kpc})$, where we cannot trace any emission line. 
However, we exclude that the extended emission could belong to another object, because its redshift is compatible with the rotation curve of the galaxy derived from the central region of the spectrum.
Moreover, both optical and radio images do not show any close object bright  enough to produce such an emission.
The apparent separation is probably due to the lack of gas in that region of the galaxy. 
Such kind of structure is not observed neither in NGC\,7212 nor in IC\,5063.
Deep images with narrow-band filters are needed to understand the complete morphology of the emission.

An exam of the H$\beta$ region of the spectrum extracted at $27\,\si{arcsec}$ from the nucleus (Fig.\,\ref{fig:mrk783}, panel C), seems to indicate that the [O III]/H$\beta$ ratio is higher with respect to what expected for star forming regions.
Therefore, a preliminary diagnostic diagram \citep{Baldwin81,Veilleux87} was assembled (Fig.\,\ref{fig:diag783}, right panel) using only the narrow component of the broad lines (H$\alpha$ and H$\beta$) and binning the spectrum in $3$ px bins in the spatial direction to increase the signal-to-noise ratio (SNR).
In some cases we could not measure some of the emission lines needed to the plot (typically H$\beta$ and [N II]), so we estimated an upper limit for their flux using the rms of the continuum and the FWHM of the [O III] line.
We used \citet{Kewley06} relation to discern the ionization mechanism of the gas.
The diagram in Fig.\,\ref{fig:diag783} (right panel) shows that most of the extended emission is photo-ionized by the AGN, particularly the most extended part, while closer to the nucleus there might be contamination by star formation.

Interestingly, star contamination is excluded in both the other galaxies presented in this paper.
Additional higher quality observations are needed to study the ENLR in NLS1 in general and Mrk 783 in details.

\section{Summary}
\label{sec:summ}

We  report the main results of our recent work on the ENLR of nearby AGN.
We used high resolution spectra of two nearby Seyfert 2 galaxies, together with combined models of photo-ionization and shocks to study the properties of the gas, in general, and  as  function of velocity in the different galaxy regions, in particular.
We found several evidences of interaction between the ISM of the galaxies and their radio jets, such as a) the contribution of shocks in ionizing the high velocity gas, b) the complex kinematics showed by the profile of the emission lines, c) the high fragmentation of matter, etc. 
The results show that the ENLR of IC\,5063 most likely has a hollow bi-conical shape, with one edge aligned with the galaxy disk. 
This may explain the velocity dependence of the ionization parameter observed in this galaxy.

We then compared the ENLR of our two galaxies with the preliminary results of the analysis of a newly discovered ENLR in a NLS1 galaxy: Mrk 783.
The object shows an extended optical emission aligned with the radio emission first observed by \citet{Congiu17}.
The extension of the optical emission is almost twice the size of the radio emission which seems mainly due to the AGN activity, even though there is contamination by star formation around $12$ arcsec from the nucleus.
While star contamination was excluded by the diagnostic diagrams both in IC\,5963 and NGC\,7212 and shock contribution was used to explain the spectra of high velocity gas, it seems that in Mrk 783 there might be a significant contribution of star formation to the extended optical emission.

Further detailed observations of ENLRs will be crucial to investigate the still poorly understood feedback between AGN and host galaxy, and to understand the effects of relativistic jets on their surrounding environment.

\section*{Author Contributions}

EC was responsible for most of the data reduction, data analysis and the text. MC produced all the simulations and contributed to part of the text. SC and VC contributed to the main idea of the paper and the data analysis. FDM observed all the target with the Magellan telescopes and with the review of the final text. MB, MF, GLM and PR contributed with the data reduction and analysis and with the review of the final text.

\section*{Acknowledgments}
This research has made use of the NASA/IPAC Extragalactic Database (NED) which is operated by the Jet Propulsion Laboratory, California Institute of Technology, under contract with the National Aeronautics and Space Administration. 
This paper includes data gathered with the 6.5-m Magellan Telescopes located at Las Campanas Observatory, Chile. 
This paper is based on observations collected with the 1.22m Galileo telescope of the Asiago Astrophysical Observatory, operated by the Department  of Physics and Astronomy "G. Galilei" of the University of Padova.
Funding for the Sloan Digital Sky Survey has been provided by the Alfred P. Sloan Foundation, and the U.S. Department of Energy Office of Science. 
The SDSS web site is \url{http://www.sdss.org}. 
SDSS-III is managed by the Astrophysical Research Consortium for the Participating Institutions of the SDSS-III Collaboration including the University of Arizona, the Brazilian Participation Group, Brookhaven National Laboratory, Carnegie Mellon University, University of Florida, the French Participation Group, the German Participation Group, Harvard University, the Instituto de Astrofisica de Canarias, the Michigan State/Notre Dame/JINA Participation Group, Johns Hopkins University, Lawrence Berkeley National Laboratory, Max Planck Institute for Astrophysics, Max Planck Institute for Extraterrestrial Physics, New Mexico State University, University of Portsmouth, Princeton University, the Spanish Participation Group, University of Tokyo, University of Utah, Vanderbilt University, University of Virginia, University of Washington, and Yale University.

\bibliographystyle{frontiersinSCNS_ENG_HUMS} % for Science, Engineering and Humanities and Social Sciences articles, for Humanities and Social Sciences articles please include page numbers in the in-text citations
\bibliography{bibliografia_def}

%%% Make sure to upload the bib file along with the tex file and PDF
%%% Please see the test.bib file for some examples of references

\section*{Figure captions}

%%% Use this if adding the figures directly in the mansucript, if so, please remember to also upload the files when submitting your article
%%% There is no need for adding the file termination, as long as you indicate where the file is saved. In the examples below the files (logo1.jpg and logo2.eps) are in the Frontiers LaTeX folder
%%% If using *.tif files convert them to .jpg or .png
%%% If using panelled/compiled figures with subcaptions, use the subcaption package as in the second example.  N.B. This package is incompatible with the subfigure and subfig packages
%%%  NB logo1.jpg is required in the path in order to correctly compile front page header %%%

%%% If you don't add the figures in the LaTeX files, please upload them when submitting the article.

%%% Frontiers will add the figures at the end of the provisional pdf automatically %%%

%%% The use of LaTeX coding to draw Diagrams/Figures/Structures should be avoided. They should be external callouts including graphics.

\end{document}